\documentclass[10pt,prb,aps,twocolumn,showpacs,superscriptaddress]{revtex4}

\usepackage{graphicx}
\newcommand{\be}{\begin{equation}}
\newcommand{\ee}{\end{equation}}
\newcommand{\bea}{\begin{eqnarray}}
\newcommand{\eea}{\end{eqnarray}}

\begin{document}

\title{Criticality in coupled quantum spin-chains with competing
ladder-like and two-dimensional couplings}

\author{Pinaki Sengupta}
\altaffiliation{Present address: Department of Physics, University of California,
Riverside, California 92521}
\email{pinaki@physics.ucr.edu}
\affiliation{Department of Physics, University of California,
Davis, California 95616}

\author{Weihong Zheng}
\email{w.zheng@unsw.edu.au}
\affiliation{School of Physics,
 University of New South Wales,
 Sydney, NSW 2052, Australia}

\author{Rajiv R. P. Singh}
\email{singh@physics.ucdavis.edu}
\affiliation{Department of Physics, University of California,
Davis, California 95616}
\date{\today}

\begin{abstract}
Motivated by the geometry of spins in the material CaCu$_2$O$_3$,
we study a two-layer, spin-half Heisenberg model, with
nearest-neighbor exchange couplings $J$ and $\alpha J$ along the two
axes in the plane and a coupling $J_\perp$ perpendicular
to the planes. We study these class of models using
the Stochastic Series Expansion (SSE) Quantum
Monte Carlo simulations at finite temperatures and series expansion
methods at $T=0$. The critical value of the interlayer coupling,
$J_\perp^c$, separating the N{\'e}el ordered and disordered ground states,
is found to follow very closely a square root dependence
on $\alpha$. Both $T=0$ and finite-temperature properties of
the model are presented.
\end{abstract}


\maketitle

\section{Introduction}

In recent years, the geometrical arrangement of atoms carrying
spin and those mediating interactions between them, in newly
discovered magnetic materials, has been a dominant theme in
the field of quantum magnetism, and has inspired many interesting
models and theories \cite{cmp}. In a class of cuprate materials, the
arrangement of copper and oxygen atoms in the copper-oxide planes
leads to nearly ideal realizations of 1D spin-chains, of spin-ladders
as well as of simple and decorated square-lattice
systems \cite{dagotto, kim}. The
variations in geometry, in other recently discovered materials, have led
to several magnetically ordered, spin-gap and dimerized phases \cite{srcb}. These
materials have allowed the exploration of dimensional crossovers,
unusual excitations and quantum critical phenomena \cite{CHN,CSY} through detailed
comparisons between experiments and theory.

Many classes of materials involve weakly coupled spin-half chains.
From a theoretical point of view, these are ideal systems for finding and
testing novel phenomena as our understanding of the spin-half chain is rather
complete thanks to Bethe ansatz methods, quantum field-theory methods
and numerical techniques \cite{affleck,tsvelik,white}.
However, different ways of coupling the
chains together can lead to widely different behaviors. If two Heisenberg
chains are coupled weakly
together, it is known to lead to a disordered ground state
and a gap in the spin-excitation spectra. These systems have been called
spin-ladders and are considered generic
examples of spin-gap phenomena \cite{dagotto,oitmaa}.
On the other hand, if an array of spin-chains, say arranged parallel
to each other in a  two-dimensional plane,
are weakly coupled together in an unfrustrated manner, they develop conventional
N{\'e}el order \cite{affleck1}.
If the inter-chain couplings are strongly frustrated, one can
even find a carryover of exotic one-dimensional
physics to higher dimensions \cite{starykh,tsvelik1}.

In this paper, we are motivated by the material CaCu$_2$O$_3$, which has
a geometry very similar to the spin-ladder material SrCu$_2$O$_3$,
but develops N{\'e}el order at low temperatures \cite{kiryukhin}.
The largest exchange coupling in these materials is of order $2000K$ and
leads to spin-half chains. These chains are weakly coupled in
two-different ways. Along one axis perpendicular to the chains, one
has effectively a two-leg ladder, which are then nearly decoupled
from the rest of the system, due to the geometry of copper-oxygen
bond angles. Along the other axis perpendicular to
the chains, the coupling between the chains leads to a square-lattice
of spins \cite{rosner}. Thus the material has competing tendencies to N{\'e}el order
and to develop a gap in the spin excitation spectra. N{\'e}el
order is found \cite{kiryukhin} in bulk materials with $T_N=25K$.
In a Science
paper, Sch{\"o}n et. al. \cite{schon} had argued that thin films of the material act
as spin-ladders and show field-induced superconducitvity. While controversy
surrounding the work of Sch{\"o}n et. al. puts the existence of field-induced
superconductivity in doubt, the fact remains that
this material is very close to being quantum critical and could go
into an ordered or disordered phase with small changes in parameters.

Here we consider two square-lattice layers of spins, with axes of the
square-lattice pointing along x and y and
a Heisenberg Hamiltonian
\begin{equation}
{\cal H}=J\sum_{a,i} S_{a,i} S_{a,i+\hat{x}} +\alpha J\sum_{a,i} S_{a,i} S_{a,i+\hat{y}}
+ J_\perp\sum_{i} S_{1,i} S_{2,i} \label{eq_H}
\end{equation}
Here, $a$ takes values $1$ and $2$,
the first sum runs over the nearest-neighbors along the x-axis,
the second over the nearest neighbor along the y-axis and the third
between the nearest-neighbors along the z-axis. The model ${\cal H}$,
in the limit of $\alpha=1$ (the isotropic bilayer), has been extensively
studied using numerical and analytical methods.\cite{previous} We shall
call ${\cal H}$
an anisotropic bilayer model. We set $J=1$ and study the phase diagram
of this model at $T=0$ using Series Expansion Methods
and Stochastic Series Expansion (SSE)
Quantum Monte Carlo (QMC) simulations. We find that
the phase boundary separating the N{\'e}el ordered and disordered phases
follows very closely the behavior
\begin{equation}
J_\perp^c= A \sqrt{\alpha},
\end{equation}
where $A\approx 2.53$ is the critical point for isotropic case.
We also calculate the uniform susceptibility and the internal energy
of the model at finite temperatures, and study its
excitation spectra in the spin-gap and N{\'e}el ordered phases.

The plan of the paper is as follows: In section II, we discuss the
series expansion method. In section III, we discuss the stochastic
series expansion calculations. In section IV, we present the $T=0$
properties of the system. In section V,
we present the uniform susceptibility and internal energy
at finite temperatures and in
section VI, we present our conclusions.

\section{Series Expansion Methods}

We have carried out dimer expansions and Ising expansions for this system
at $T=0$. The linked-cluster series expansion method
has been previously reviewed in Ref. \onlinecite{gel00},
and will not be repeated here.

\subsection{Dimer Expansions}
In the limit that the exchange coupling along the rung $J_\perp$
is much larger than the couplings  within the plane,
 the rungs interact only weakly
with each other, and the dominant configuration
in the ground state is the product state with the spin
on each rung forming a spin singlet.
We can construct dimer expansion in $J/J_\perp$ by treating
the last term in Eq. (\ref{eq_H}) as the unperturbed Hamiltonian
and the rest of terms as  a perturbation.

We have carried out the dimer expansions for the $T=0$ ground state energy
per site, $E_0/N$,  the antiferromagnetic susceptibility $\chi$,
and the lowest lying triplet excitation spectrum $\Delta (k_x, k_y)$
(odd parity under interchange of the planes)
up to order $(J/J_\perp)^{11}$ for fixed values of $\alpha$.
The resulting power series in $J/J_\perp$ for the ground state energy
per site $E_0/N$ and the antiferromagnetic susceptibility $\chi$
for $\alpha=1/2$ are presented in Table I.
A table of series coefficients for the triplet excitation
spectrum $\Delta (k_x, k_y)$ would require an inordinate amount
of space to reproduce in print; it is available from the authors
upon request. Instead, we also present in Table I the series
for the minimum energy gap $m=\Delta (\pi, \pi)$.
The dimer expansion for isotropic case ($\alpha=1$) was  carried out
by Hida\cite{hid} up to order  $(J/J_\perp)^{6}$
in 1992, and  extended to order $(J/J_\perp)^{8}$ by Gelfand\cite{gel},
and to order $(J/J_\perp)^{11}$ by one of the authors\cite{zwh}.
Here, the number of clusters involved is much more than the isotropic case
 since the system no longer has 90$^o$ rotation symmetry, and there are,
in all, 38070 linked clusters of up to 12 sites involved in the calculation.

\subsection{Ising Expansions}
To construct an expansion about the Ising limit for this system,
one has to introduce an anisotropy parameter $x$,
and write the Hamiltonian for Heisenberg-Ising model as
\begin{equation}
H = H_0 + x V~,  \label{Hising}
\end{equation}
where

\begin{eqnarray}
H_0 &= & J \sum_{a=1,2} \sum_{i} S_{a,i}^z S_{a,i+\hat{x}}^z
+ \alpha J \sum_{a=1,2} \sum_{i} S_{a,i}^z S_{a,i+\hat{y}}^z \nonumber \\
&& + J_\perp \sum_{i} S_{1,i}^z S_{2,i}^z +
 t \sum_{\alpha,i} \epsilon_{\alpha,i} S_{\alpha,i}^z~, \nonumber \\
V &= &
J \sum_{\alpha=1,2} \sum_{i} ( S_{\alpha,i}^x S_{\alpha,i+\hat{x}}^x
        + S_{\alpha,i}^y S_{\alpha,i+\hat{x}}^y ) \nonumber \\
&& + \alpha J \sum_{\alpha=1,2} \sum_{i} ( S_{\alpha,i}^x S_{\alpha,i+\hat{y}}^x
        + S_{\alpha,i}^y S_{\alpha,i+\hat{y}}^y ) \nonumber \\
&& + J_\perp \sum_{i} ( S_{1,i}^x S_{2,i}^x + S_{1,i}^y S_{2,i}^y )
- t \sum_{\alpha,i} \epsilon_{\alpha,i} S_{\alpha,i}^z ~,
\end{eqnarray}

and $\epsilon_{\alpha,i}=\pm 1 $ on the two sublattices. The last term
in both $H_0$ and $V$ is a local staggered field term, which can be
included to improve convergence.
The limits $x=0$ and $x=1$ correspond to the Ising model and
the isotropic Heisenberg model, respectively.
The operator $H_0$ is taken as the unperturbed
Hamiltonian, with the unperturbed ground state being the
usual N\'{e}el state.
The operator $V$ is treated as  a perturbation.
It flips a pair of spins on neighboring sites.

The Ising series have been calculated for
the ground state energy per site, $E_0/N$,
the staggered magnetization $M$,  the uniform
perpendicular susceptibility $\chi_{\perp}$,
and the lowest lying triplet excitation spectrum $\Delta (k_x, k_y, k_z)$
for several ratio of couplings and (simultaneously) for several values of $t$
 up to order $x^{10}$
(the series for uniform perpendicular susceptibility $\chi_{\perp}$
 is one order less).
The series are
available upon request.
Here there are two branches of the spin-wave dispersion.
From the series
one can see
that the symmetric (optical) excitation spectrum  $\Delta (k_x, k_y, 0)$
is related to the antisymmetric
(acoustic) excitation spectrum $\Delta (k_x, k_y, \pi)$
by
\begin{equation}
\Delta (k_x, k_y, 0)= \Delta (\pi-k_y, \pi-k_x, \pi)~,
\end{equation}
and so we only consider the antisymmetric excitation spectrum here.

\section{Quantum Monte Carlo simulations}

We have used the stochastic series expansion (SSE) method \cite{sse1,sse2}
to study the ground state and finite temperature properties of the Heisenberg
antiferromagnet on anisotropic bilayers. The SSE is a
finite-temperature QMC technique based on importance sampling of the diagonal
matrix elements of the density matrix $e^{-\beta H}$. Ground state properties
are obtained by using sufficiently large values of $\beta$. There are no
approximations beyond statistical errors. Using the ``operator-loop''
cluster update,\cite{sse2} the autocorrelation time for the system sizes
we consider here (up to $\approx \times 10^3$ spins) is at most a few Monte
Carlo sweeps even at the critical coupling.\cite{dloops}

\section{$T=0$ properties}

\begin{figure}
\includegraphics[width=8.3cm]{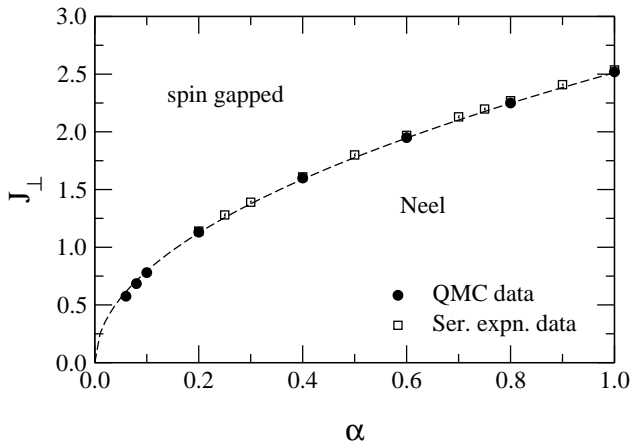}
\caption{The ground state phase diagram in the parameter space of the in-plane
anisotropy, $\alpha$, and the inter-layer coupling, $J_{\perp}$. Both QMC
and series expansion results are shown. The error bars are smaller than
the size of the symbols. The curve is a square root fit to the critical coupling data.}
\label{fig:phase}
\end{figure}

\subsection{Results from Dimer Expansions}

With the dimer series for the antiferromagnetic susceptibility $\chi$,
and the minimum triplet gap $m$, one can
determine the critical point $(J/J_{\perp})_c$ by constructing Dlog Pad\'{e}
approximants\cite{gut} to these series,
and since the transition should lie in
the universality class of the classical $d=3$ Heisenberg model
(our unbiased analysis also supports this), we expect that the critical
index for $\chi$ and $m$
should be approximately  1.40 and 0.71, respectively.
The critical line obtained by the
exponent-biased Dlog Pad\'{e} approximants\cite{gel}
are shown in Fig. \ref{fig:phase}.

\begin{figure}
\includegraphics[width=8.3cm]{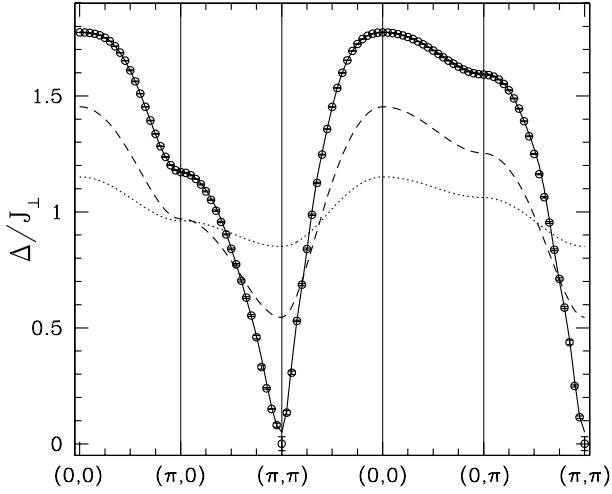}
\caption{
Plot of the antisymmetric spin-triplet excitation spectrum
$\Delta (k_x, k_y, \pi)/J_{\perp}$ in dimer phase
 along high-symmetry cuts through the Brillouin
zone for the system with coupling ratios
$\alpha=0.5$ and $J/J_\perp=0.1$ (dotted line), 0.3(dashed line), 0.556 (solid line).
The lines are the estimates by direct sum to the
dimer series, and the points circles with error bar for the case of
$J/J_\perp=0.556$ only are the estimates of the integrated differential  approximants
 to the dimer series.
}
\label{fig:mk_dimer}
\end{figure}

\begin{figure}
\includegraphics[width=8.3cm]{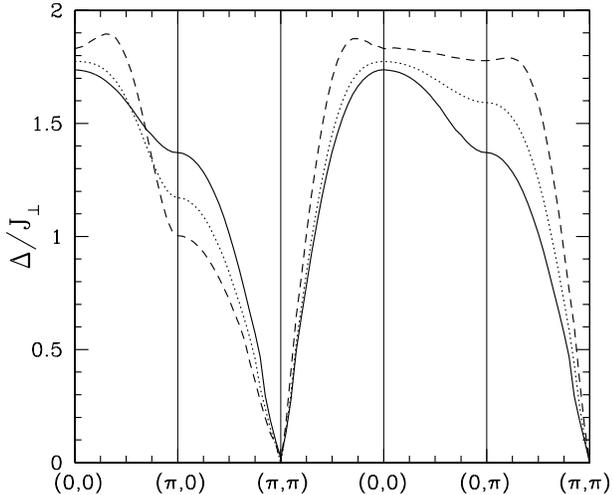}
\caption{
Plot of the antisymmetric spin-triplet excitation spectrum
$\Delta (k_x, k_y, \pi)/J_{\perp}$
 along high-symmetry cuts through the Brillouin
zone for the system along the critical line with
$\alpha=0.25$ (dashed line), 0.5 (dotted line), 1(solid line).
The results of the integrated differential  approximants to
the dimer series are shown.
}
\label{fig:mk_xc}
\end{figure}

The spectra for some particular values of $\alpha$ and $J/J_\perp$ in
the dimer phase are
illustrated in Fig. \ref{fig:mk_dimer},
where the direct sum to the series at $J/J_\perp=(J/J_\perp)_c$ is indeed consistent
with the integrated differential approximants\cite{gel} that one can construct.
In Fig. \ref{fig:mk_xc}, we show the spectra for some particular values of $\alpha$ and
$J/J_\perp$ along the critical line.

\begin{figure}
\includegraphics[width=8.3cm]{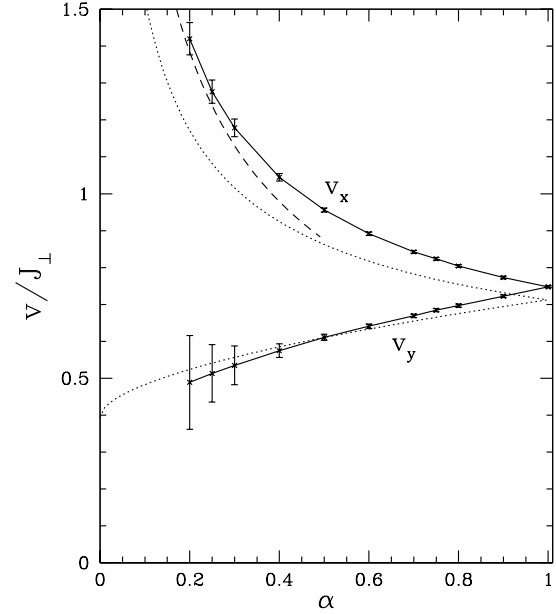}
\caption{
The spin-wave velocity $v/J_{\perp}$ along x and y directions
versus $\alpha$ at the critical ratio $(J/J_{\perp})_c$
obtained from dimer series expansion. Also shown are the asymptotic results of
$v_x/J_\perp=0.619/\sqrt{\alpha}$ (dashed line) and
the linear spin-wave results for $v_x$ and $v_y$ (dotted lines).
}
\label{fig:v_xc}
\end{figure}

To compute the critical spin-wave velocity $v$, one expands the spectrum
$\Delta$ in the vicinity of wave-vector $(\pi,\pi)$ up to ${\bf k}^2$:
\bea
&& \Delta (\pi-k_x, \pi-k_y)(J/J_\perp) = C(J/J_\perp) \nonumber \\
&& \quad + D_x(J/J_\perp) k_x^2 + D_y(J/J_\perp) k_y^2 + \cdots~,
\label{eqv}
\eea
and it is easy to prove\cite{gel} that the critical spin-wave velocity along x and y
 direction are
equal to $(2C D_x)^{1/2}$ and $(2C D_y)^{1/2}$, respectively, at $(J/J_\perp)_c$.
The series coefficients for $2CD_x$ and $2CD_y$ in $J/J_\perp$ are listed in Table I.
These series can be extrapolated to $(J/J_\perp)_c$ by using the integrated differential
approximants, and the results are shown in Fig. \ref{fig:v_xc}, one can see that
 $v_x$ ($v_y$) is increased (decreased) once the anisotropy is introduced.
As $\alpha\to 0$, the system approaches a spin-chain along the x-axis,
where the spin-wave velocity is known to be $\pi J/2$.
Hence asymptotically, along the critical line,
\begin{equation}
v_x/J_\perp\simeq (\pi/2)(J/J_\perp)\sim 0.619/\sqrt{\alpha}
\end{equation}
This asymptotic behavior is shown by a dashed line. One can see that
the presence of other interactions further
increases the spin-wave velocity along x.
To understand the behavior of $v_y$, we turn to linear spin-wave theory.
We obtain the antisymmetric excitation spectrum
\bea
&& \Delta(k_x, k_y) = 2 S J {\Big [} {\big (} 1 + \alpha + J_\perp/2J {\big )}^2 \nonumber \\
&& \quad - {\big (}  \cos ( k_x) + \alpha \cos (k_y) - J_\perp/2J {\big )}^2 {\Big ]}^{1/2}
\eea
with this one can obtain spin-wave velocity along the y-direction
\begin{equation}
v_y=2SJ [\alpha (1+\alpha+J_\perp/J)]^{1/2}.
\end{equation}
Note that $v_x$ is obtained from $v_y$ by exchanging the
intralayer couplings along x and y-directions.
The linear spin-wave results for both $v_x$ and $v_y$
are plotted as dotted lines in Fig. \ref{fig:v_xc}.
We see that for small $\alpha$, the linear spin-wave theory, which is known
to be off by a factor of $\pi/2$ for the spin-half chain, works quite
well for $v_y$, the velocity along the weakly-coupled direction.

\subsection{Results from Ising Expansions}

\begin{figure}
\includegraphics[width=8.3cm]{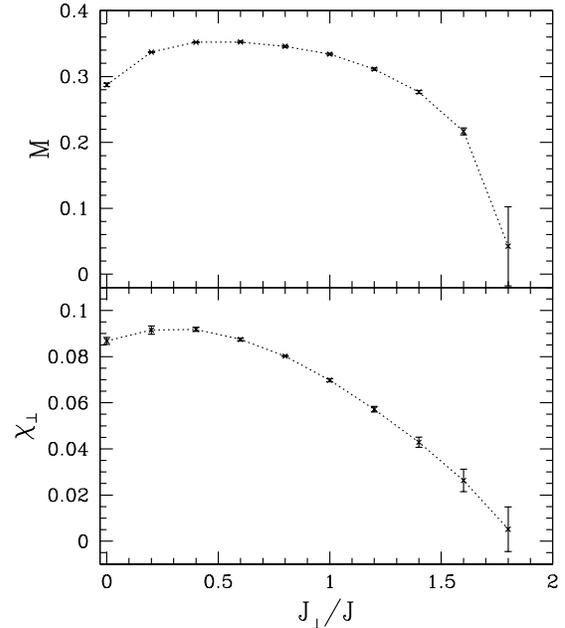}
\caption{
The staggered magnetization $M$ and the  uniform perpendicular susceptibility $\chi_\perp$
 vs $J_\perp/J$ for $\alpha=0.5$ as estimated by Ising expansions.
}
\label{fig:M_chi}
\end{figure}

With the Ising series for staggered magnetization $M$ and the uniform
perpendicular susceptibility $\chi_{\perp}$, one can also determine the
phase boundary by extrapolating the series for $M$ and $\chi_\perp$
to the isotropic point $x=1$
using the same method as in Ref. \onlinecite{zwh}, the results for $\alpha=1/2$
are shown in Fig. \ref{fig:M_chi},
We note that $M$ and $\chi_{\perp}$  first increase
for small $J_\perp/J$,
then decrease for a larger value of $J_\perp/J$,
 and vanish at about $J_\perp/J=1.8$, which is consistent
with the more accurate critical point determined by the dimer expansions.
The reason for the initial increase is that for small $J_\perp/J$
the interlayer coupling enhances the antiferromagnetic long-range order
as the system acquires a weak three dimensionality and
quantum fluctuations are suppressed.

\begin{figure}
\includegraphics[width=8.3cm]{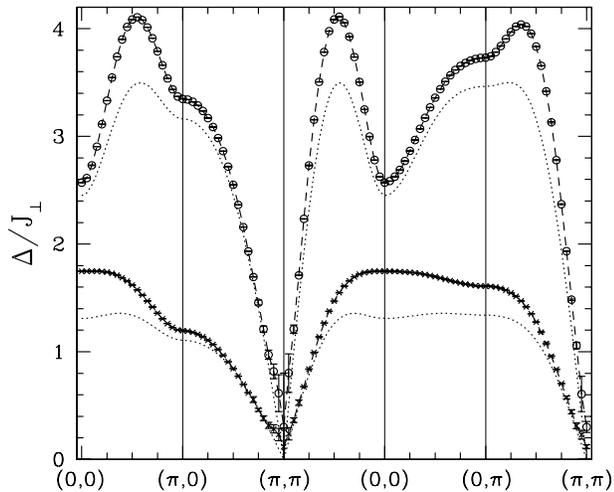}
\caption{
Plot of the antisymmetric spin-triplet excitation spectrum
$\Delta (k_x, k_y, \pi)/J_{\perp}$ in N{\'e}el order phase
derived from the Ising expansions along high symmetry
cuts through the Brillouin zone for the system with coupling
ratios $\alpha=0.5$ and $J_\perp/J=0.5$ (upper curve), 1.75 (lower curve),
also shown as dotted lines are the results of linear spin-wave theory.
}
\label{fig:mk_neel}
\end{figure}

The antisymmetric excitation spectra $\Delta (k_x, k_y, \pi)$
for some particular values of $J_\perp/J$ and $\alpha=1/2$ are illustrated in
Fig. \ref{fig:mk_neel}, where we can see that the excitation is gapless
at the  $(\pi,\pi,\pi)$ point.
For $J_\perp/J=1.75$ (close to the critical point), the spectra are very similar to that obtained from dimer
expansion. Thus, as one goes through the quantum phase transition, the
spectra evolve smoothly.

\subsection{Ground State Properties Obtained by SSE}

\begin{figure}
\includegraphics[width=8.3cm]{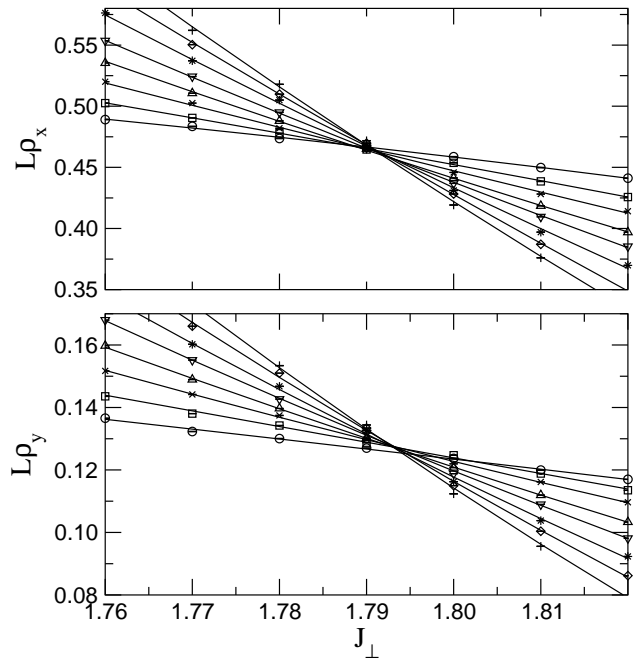}
\caption{The ground state spin stiffness times the linear system size, $L$,
as a function of the inter-layer coupling in the vicinity of the critical
point for square lattices with $L=6,8,\cdots,20$. The (negative) slope
increases with increasing $L$. The intra-layer anisotropy is $\alpha=0.5$.
The upper (lower) panel shows the data for stiffness along
the x-(y-) axis. Error bars are of the size of the symbols or smaller.The
curves are quadratic fits to the data.}
\label{fig:rhoxy}
\end{figure}

\begin{figure}
\includegraphics[width=8.3cm]{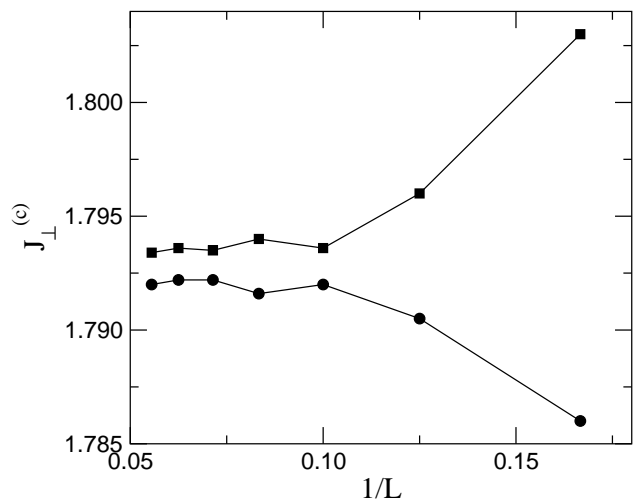}
\caption{The finite-size dependent critical point as obtained from the
crossing of the stiffness curves for successive system sizes with linear
dimensions $L$ and $L+2$. The estimate for the critical coupling converges
from above (below) for the $\rho_y (\rho_x)$ data, providing upper and lower
bounds for the true critical coupling in the thermodynamic limit.}
\label{fig:j3c}
\end{figure}

For the ground state properties, using the SSE method,
we have studied lattices of
the form $2\times L\times L$, with $L$ up to 20. Periodic boundary conditions
were applied in both the x- and y-directions. A series of values of $\alpha$
were chosen, and the critical $J_\perp$ was determined for each of them,
mapping out the ground state phase diagram in the $\alpha-J_\perp$
parameter space.
An inverse temperature $\beta=16L$ was found to be sufficient for the
calculated quantities to have converged to their ground state values.

An efficient way to determine the critical coupling for the spin gap
transition is by studying the finite size scaling of the ground state
($T$=0) spin-stiffness. The spin stiffness can be defined
\cite{kohn,kopietz} as the second derivative of the free energy with
respect to a uniform twist $\phi$. At $T=0$, the free energy is the
same as the internal energy, and the expression for the spin stiffness
takes the form
\begin{equation}
\rho = \frac{\partial ^2E(\phi)}{\partial\phi ^2},
\label{eq:rhoc}
\end{equation}
where $E(\phi)$ is the internal energy per spin in the presence of a
twist. The stiffness can be related to the fluctuations of the
``winding number'' in the simulations \cite{pollock,harada,sse1,cuccoli}
and hence can be estimated directly without actually including a twist.
Since the twist can be applied parallel to the x- or y-axis, there are two
different spin stiffnesses, $\rho_x$ and $\rho_y$, in the anisotropic
system considered here.

Finite size scaling analysis dictates that at the critical coupling,
the spin stiffness should scale with the system size as
\begin{equation}
\rho(L)\sim L^{d-2-z},
\end{equation}
where $d(=2)$ is the dimensionality and $z$ is the dynamical critical
exponent. The transition is expected to be in the universality class
of the 3D Heisenberg model -- hence $z=1$. It follows from the above
relation that in a plot of $L\rho_{\alpha}, (\alpha=x,y)$ versus $J_{\perp}$,
the curves for different system size should intersect at $J_{\perp}^c$. In
practice, it is found that the crossing point shifts monotonically with
increasing system size -- the intersection of the curves for successive
system sizes (linear dimensions $L$ and $L+2$) give a finite size
dependent estimate of the critical coupling, $J_{\perp}^c(L)$, that converges
to the true critical coupling, $J_{\perp}^c$, at large $L$. Interestingly, the
convergence is from opposite directions for $\rho_x$ and $\rho_y$ --
$J_{\perp}^c(L) \rightarrow J_{\perp}^c$ from above (below) for
$\rho_y (\rho_x)$.\cite{anders}
This yields upper and lower bounds for the true critical coupling,
leading to an improved estimate for $J_{\perp}^c$. The results
for $\alpha$=0.5 are shown in fig.\ref{fig:rhoxy}. The upper (lower) panel
shows the data for $\rho_x (\rho_y)$. The curves are found to cross in
the neighborhood of $J_{\perp} \approx 1.8$. A plot of the finite size dependent
critical coupling obtained from the crossing of successive system sizes
obtained from the plot is shown in fig.\ref{fig:j3c}. As discussed above,
$J_{\perp}^c(L)$ is seen to converge toward $\approx 1.79$ from above (below)
for $\rho_x (\rho_y)$. From the data, we estimate the true critical
coupling in the limit of infinite system size to be $J_{\perp}^c=1.79\pm 0.005$
for $\alpha=0.5$.

\begin{figure}
\includegraphics[width=8.3cm]{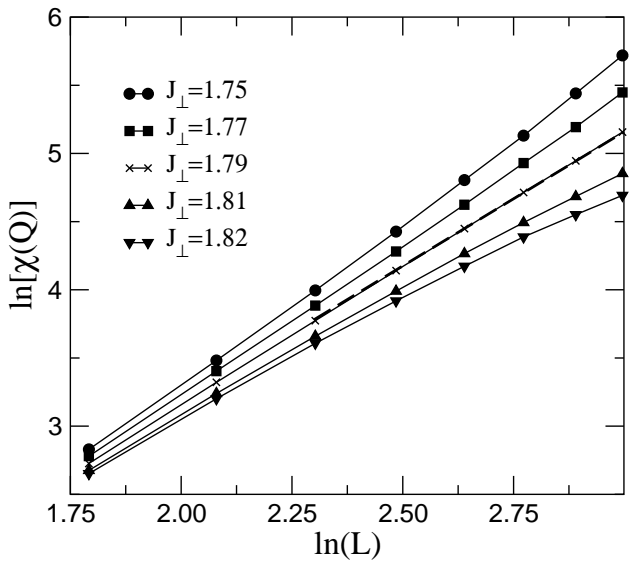}
\caption{The finite-size scaling of the staggered, {\bf Q}=$(\pi,\pi)$, static
susceptibility in the vicinity of the critical coupling for $\alpha$=0.5. For
lattices larger than $L$=10, the data close to the critical coupling is found
to fit a power law behavior with an exponent $2-\eta$ ($\eta=0.035$) that is
consistent with the 3D Heisenberg universality class.}
\label{fig:xpi}
\end{figure}

A similar analysis of the spin stiffness data for different values of
$\alpha$ gives us the critical value of the inter-planar coupling $J_\perp^c$
for the spin gap transition as a function of $\alpha$. The resulting phase
diagram is shown in fig.\ref{fig:phase}.

Independent estimate of the critical coupling can be obtained from the
finite size scaling of the staggered structure factor and the corresponding
susceptibility. The full two-plane static spin susceptibility is defined as
\begin{eqnarray}
\chi({\bf q})&=&{1\over L^2}\sum_{i,j}e^{i{\bf q}\cdot ({\bf r}_i-{\bf r}_j)}\int_0^{\beta}d\tau \langle\bigl[S^z_{1,j}(\tau)-S^z_{2,j}(\tau)\bigr]\\ \nonumber
 & &\bigl[S^z_{1,i}(0)-S^z_{2,i}(0)\bigr]
\end{eqnarray}
At the critical point, the staggered, ${\bf q}=(\pi,\pi)$, susceptibility
for finite size systems scale with the system size as a simple power law
behavior determined by the critical exponent $\eta$:
\begin{equation}
\chi(L,J_{\perp}^c)\sim L^{2-\eta}
\end{equation}
On a plot of ln($\chi$) versus ln($L$), the data should fall on a straight
line with slope $2-\eta$. In the spin gapped phase, the staggered
susceptibility should go to a constant at large $L$, whereas in the N{\'e}el
phase with long range antiferromagnetic order, it should diverge faster
than exponentially with $L$. Fig.\ref{fig:xpi} shows the plot of
ln($\chi$) versus ln($L$) for fixed $\alpha(=0.5)$ and five different values
of $J_{\perp}$ in the vicinity of the critical coupling. For large values of
$L$, the data for the
critical coupling is indeed found to fall on a straight line. The slope
of the line yields $\eta\approx 0.035$, in close agreement with its value
for the 3D Heisenberg universality class. The deviation from power law
behavior in both the spin gapped and N{\'e}el phases are in agreement with
the above discussion. Thus the transition is found to belong to the
universality class of the 3D Heisenberg model. This is found to be true
for any finite value of $\alpha$.

\begin{figure}
\includegraphics[width=8.3cm]{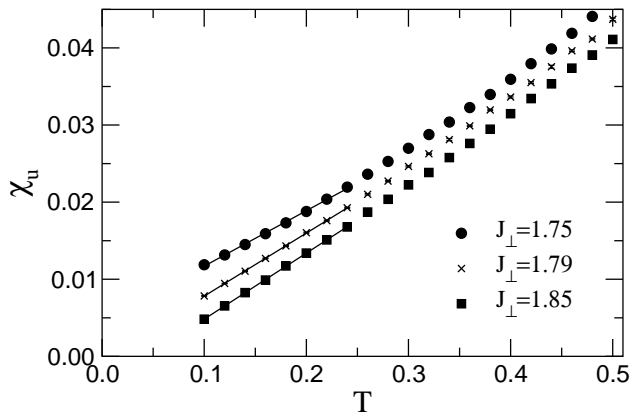}
\caption{The uniform susceptibility as a function of temperature near the
critical coupling for $\alpha$=0.5. At low temperatures, the susceptibility
depends linearly on $T$, in agreement with the prediction from field
theoretic calculations. Above the transition, in the spin gapped phase,
$\chi_u$ scales to zero at finite $T$, while below the transition, in the
N{\'e}el phase, $\chi_u$ scales to a finite value at $T$=0. At the critical
point, $\chi_u$ extrapolates to zero at $T$=0. The data shown are for
a rectangular lattice with $N$=128x32.}
\label{fig:xu}
\end{figure}

\section{Finite Temperature Properties}
\subsection{Uniform Susceptibility}

We start the presentation of the finite temperature properties with the
uniform magnetic susceptibility defined as
\begin{equation}
\chi_u={\beta\over N}\sum_{i,j}\bigl[S^z_{1,i}+S^z_{2,i}\bigr]\bigl[S^z_{1,j}+S^z_{2,j}\bigr]
\end{equation}
where $N$ is the size of the system. For finite temperature properties, we
have carried out the simulations on lattices with rectangular geometry,
$L_x\times L_y, L_x\neq L_y$. This is to reduce finite size effects. It was
shown by Sandvik \cite{multichain} for a system of coupled Heisenberg chains,
the finite size
effects depended monotonically for rectangular lattices whereas for
square lattices the behavior was less well behaved. While such effects
are expected for ground state properties also, our limitation in terms
of computational power has restricted us to the use of square lattices.
Fortunately, the stiffness in the two directions give upper and lower
bounds for the critical coupling, leading to a reliable estimate of
$J_{\perp}^c$. Simulations at finite temperatures require considerably less
computer power -- hence we are able to study rectangular lattices.
In particular, lattices with aspect ratio $L_x=4L_y$ have been considered.

Results from the study of the non-linear $\sigma$ model predict that at the
critical coupling, the uniform susceptibility, $\chi(T)$, is linear in $T$
(at low temperatures) with zero intercept. The region of linear $T$
dependence gives an estimate of the quantum critical regime.
The finite-size effects in the estimates of $T>0$ data decrease
rapidly with increasing system size -- the difference in the estimate of
$\chi_u$ between lattices with $L_x=64$ and 128 is of the order of the
magnitude of the error bars up to the lowest temperatures studied. Henceforth,
the data are presented for lattices with $L_x$=128.
Fig.\ref{fig:xu} show the temperature
dependence of the $\chi_u$ for three different values of the inter-planar
coupling close to the critical value. The data shown are for a rectangular
lattice with $L_x$=128. It was found that the difference in the estimates
of $\chi_u$ at low temperatures For all the values, the uniform
susceptibility is linear in $T$ over the range of temperature shown. For
$J_{\perp}^c$, the intercept is approximately zero, while it is positive (negative)
for $J_{\perp} <(>) J_{\perp}^c$. The results are consistent with the estimate of
$J_{\perp}^c$ obtained from the stiffness data. While the linear $T$ behavior at
the critical coupling is found to hold for all values of the in-plane
spatial anisotropy, its range decreases with decreasing value of $\alpha$.

\begin{figure}
\includegraphics[width=8.3cm]{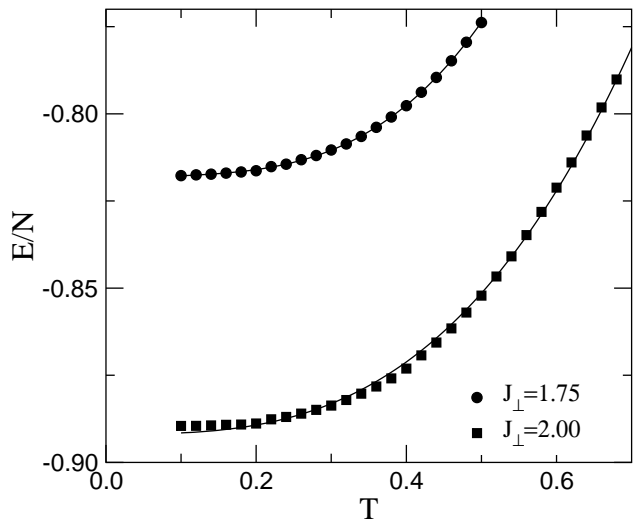}
\caption{The energy as a function of the temperature for the N{\' e}el and the
spin gapped phases. The data for $J_{\perp}$=1.85 is found to deviate from a pure
$T^3$ dependence at low temperatures.}
\label{fig:enr}
\end{figure}

\subsection{Internal Energy and Specific Heat}

Instead of working directly with the specific heat, we have studied
the internal energy, of which the specific heat is the temperature derivative.
This is driven by the practical consideration that the specific heat data
(that can be estimated directly within the framework of the SSE method) gets
noisy at low temperatures, while the internal energy data is largely
free from such noise at all temperatures considered. The temperature
dependence of the specific heat is easily obtained from that of the
internal energy. An estimate of the temperature dependence of the internal
energy is obtained by assuming spin wave dispersion at low energies.
In the N{\'e}el phase, the dispersion is $\varepsilon(k)=ck$, where $c$
is the spin wave velocity. In the presence of a spin gap ($\Delta$), the
dispersion takes the form $\varepsilon(k)=\sqrt{\Delta^2+c^2k^2}$. For a 2D
system, the internal energy per particle at low temperatures should
take the form
\[
{E\over N}\sim \int_{0}^{\infty}kdk\varepsilon(k){\frac{1}{e^{-\varepsilon(k)/k_BT} - 1}}.
\]
For the gapless dispersion, this gives ${E\over N}=(const.)\times T^3$, which
implies that the specific heat, $C_v\sim T^2$. For the spin gapped phase,
the internal energy expression reduces to
\[
{E\over N}=(const.)\times T^3 \int_{\Delta^2/T}^{\infty}{\frac{x^2dx}{e^x - 1}}.
\]
For $\Delta^2 \ll T$, the leading behavior of the specific heat is once again
$\sim T^2$. For large gap and/or low temperatures, this will turn into
an activated behavior,
coming from the temperature dependence of the definite
integral. However, this region has not been accessible to
our Monte Carlo simulations.
Fig.\ref{fig:enr} shows the internal energy as a function of the
temperature for two parameter sets corresponding to the gapless and spin
gapped phases along with the $T^3$ fit. This suggests that unless one goes
down to very low temperatures, near the critical line it would be
difficult to tell from the specific heat data whether one is in an
ordered or a spin-gap phase.

\section{Conclusions}

We have used the stochastic series expansion (SSE) quantum Monte Carlo
(QMC) and series expansion methods to study the antiferromagnetic
Heisenberg model on
spatially anisotropic bilayer systems.
The critical $J_\perp^c$,
separating the N{\'e}el ordered and disordered phases is
found to depend on $\alpha$, the ratio of in-plane couplings,
according to a simple square-root behavior.
For all values of $\alpha$ considered,
the transition to the spin-gapped state belongs to the
universality class of the 3D classical Heisenberg model.
The $T=0$ and finite temperature properties of the model
are studied, especially around the critical line. We find that the
spin-wave spectra evolve smoothly through the transition. The
spin-wave velocity becomes highly anisotropic as $\alpha\to 0$.
We hope our work would stimulate further measurements of
spin-wave spectra for materials such as CaCu$_2$O$_3$.

\acknowledgments{
We would like to thank H. Rosner, A. Sandvik,  J. Oitmaa, and
C.J. Hamer for fruitful
discussions. This work was supported in part by the US National
Science Foundation grant numbers DMR-9986948
and DMR-0240918 and by the Australian Research Council.
Part of the simulations were carried out on the
IBM SP facility at NERSC.
We are also grateful for the computing resources provided
by the Australian Partnership for Advanced Computing (APAC)
National Facility.
}

\begin{table*}
\caption{
Series coefficients for dimer expansions of the ground-state energy per site,
$E_0/NJ_\perp$, the antiferromagnet susceptibility
$\chi_{\perp}$, the minimum triplet energy gap $m/J_\perp$,
and the critical spin-wave velocity along x and y-direction
(as described in the text). Coefficients of $(J/J_{\perp})^n$  for $\alpha=1/2$
up to order $n=11$ are listed.
  } \label{tab1}
\begin{ruledtabular}
\begin{tabular}{|c|l|l|l|l|l|}
 \multicolumn{1}{|c|}{$n$} &\multicolumn{1}{c|}{$E_0/J_{\perp}$}
 &\multicolumn{1}{c|}{$\chi_{\perp}$}
 &\multicolumn{1}{c|}{$m/J_\perp$}
 &\multicolumn{1}{c|}{$2CD_x/J^2_\perp$}
 &\multicolumn{1}{c|}{$2CD_y/J^2_\perp$}\\
\hline
  0 & -3.750000000$\times 10^{-1}$ & ~1.000000000      & ~1.000000000      & ~0.000000000      & ~0.000000000      \\
  1 & ~0.000000000      & ~3.000000000      & -1.500000000      & ~1.000000000      & ~5.000000000$\times 10^{-1}$ \\
  2 & -2.343750000$\times 10^{-1}$ & ~7.125000000      & ~1.250000000$\times 10^{-1}$ & ~0.000000000      & ~0.000000000      \\
  3 & -1.054687500$\times 10^{-1}$ & ~1.546875000$\times 10^{1}$ & -4.687500000$\times 10^{-1}$ & ~1.750000000      & ~5.468750000$\times 10^{-1}$ \\
  4 & -2.270507813$\times 10^{-2}$ & ~3.132617187$\times 10^{1}$ & -5.078125000$\times 10^{-1}$ & ~1.218750000      & ~1.757812500$\times 10^{-1}$ \\
  5 & ~8.184814453$\times 10^{-2}$ & ~6.153011068$\times 10^{1}$ & -1.647949219$\times 10^{-1}$ & -4.101562500$\times 10^{-1}$ & -3.376464844$\times 10^{-1}$ \\
  6 & ~9.038543701$\times 10^{-2}$ & ~1.188306387$\times 10^{2}$ & ~2.856903076$\times 10^{-1}$ & -1.158325195      & -1.558227539$\times 10^{-1}$ \\
  7 & ~1.045632362$\times 10^{-2}$ & ~2.276296512$\times 10^{2}$ & ~7.159423828$\times 10^{-2}$ & -1.660129547      & -8.045387268$\times 10^{-2}$ \\
  8 & -1.300574541$\times 10^{-1}$ & ~4.326861726$\times 10^{2}$ & -8.715667129$\times 10^{-1}$ & ~6.815385818$\times 10^{-1}$ & ~7.272934914$\times 10^{-1}$ \\
  9 & -1.566342926$\times 10^{-1}$ & ~8.165576512$\times 10^{2}$ & -2.380170342      & ~3.975312367      & ~8.960766308$\times 10^{-1}$ \\
 10 & -2.075107419$\times 10^{-2}$ & ~1.529763646$\times 10^{3}$ & -1.890960221      & ~2.580596185      & -2.333164847$\times 10^{-1}$ \\
 11 & ~2.349058435$\times 10^{-1}$ & ~2.850279307$\times 10^{3}$ & -1.916425429$\times 10^{-1}$ & -2.162085212      & -1.298405600      \\
\end{tabular}
\end{ruledtabular}
\end{table*}


\begin{thebibliography}{00}
\bibitem{anders}
We would like to thank Anders Sandvik for suggesting such a finite-size behavior
of the stiffness along the two principal axes.
\end{thebibliography}
\end{document}